\documentclass[aps,prd,twocolumn,superscriptaddress]{revtex4}

\usepackage{amsmath}
\usepackage{color}
\definecolor{bl}{cmyk}{1,1,0.3,0}
\usepackage[dvipdfm]{hyperref}
\hypersetup{colorlinks=true,bookmarks=true,urlcolor=bl,linkcolor=bl,citecolor=bl}

\begin{document}


\title{Massive gravity and structure formation}


\author{Michael V. Bebronne}
\affiliation{Service de Physique Th\'eorique, Universit\'e
Libre de Bruxelles (U.L.B.), CP225, boulevard. du Triomphe, B-1050
Bruxelles, Belgium.}
\author{Peter G. Tinyakov}
\affiliation{Service de Physique Th\'eorique, Universit\'e
Libre de Bruxelles (U.L.B.), CP225, bld. du Triomphe, B-1050
Bruxelles, Belgium.}
\affiliation{Institute for Nuclear Research of the Russian
Academy of Sciences, 60th October Anniversary
Prospect, 7a, 117312 Moscow, Russia.}


\date{\today}

\begin{abstract}
We study the growth of cosmological perturbations in the model of
Lorentz-violating massive gravity. The Friedman equation in this model
acquires an unconventional term due to the Lorentz-breaking
condensates which has the equation of state $w=-1/(3\gamma)$ with
$\gamma$ being a free parameter taking values outside of the range
$[0,1/3]$. Apart from the standard contributions, the perturbations
above the Friedmann background contain an extra piece which is
proportional to an arbitrary function $\vartheta(x^i)$ of the space
coordinates. This function appears as an integration constant and
corresponds to a non-propagating scalar mode which may, however,
become dynamical with the account of the higher-derivative
corrections. For $-1<\gamma<0$ and $\gamma=1$ the ``anomalous''
perturbations grow slower than the standard ones and thus the model is
compatible with observations. Whether the model is experimentally
acceptable at other values of $\gamma$ depends on the value of the
function $\vartheta(x^i)$ at the beginning of the radiation-dominated
epoch.
\end{abstract}

\preprint{ULB-TH/07-20}
\maketitle

\section{Introduction}

The standard cosmological model is based on the assumption that the
gravitational interaction is correctly described by the general
relativity (GR) at scales comparable to the horizon size. This model
is quite successful in describing the bulk of the cosmological data:
the anisotropies of the cosmic microwave background radiation, the
primordial abundance of light elements and the structure formation in
the early Universe. The quantitative agreement between the standard
cosmological model and the observations has an ever-growing precision
\cite{Spergel:2006hy}. The question arises to which extent one should
consider this agreement as a confirmation of the general relativity
itself.

In order to address this question an alternative model is needed whose
predictions can be compared to those of GR. Such a model should
coincide with GR at scales from $\sim 0.1$~mm to the size of the solar
system where GR has been tested directly. Therefore, the modifications
of the gravitational interaction should occur in the infrared, at
distances much larger than the size of the solar system.

Interestingly, at these distances the predictions of GR actually {\em
do not agree} with the observations; only after the introduction of
the otherwise undetected dark matter and dark energy the agreement is
achieved. The necessity for these new components is a major problem of
the standard cosmology. Hence, in parallel with the direct searches
for the dark components, the alternative models of gravity should be
explored which may eventually eliminate the need (or provide
alternative candidates) for the dark matter and shed new light on the
nature of the dark energy.

It is a challenging problem to modify the gravitational interaction at
large distances. An alternative model has to be theoretically
consistent, i.e., free from ghosts and instabilities. In addition, it
should be in agreement with the existing experimental data and should,
ideally, provide testable predictions for the future experiments.  It
is not obvious that such models exist, so it would be very
important to construct an example.

There have been several attempts made in this direction
\cite{Milgrom:1983pn,Bekenstein:2004ne,Gregory:2000jc,Dvali:2000hr,Carroll:2003wy,Kogan:2000vb,Damour:2002ws,Arkani-Hamed:2003uy,Rubakov:2004eb}.
In this paper we concentrate on the massive gravity model
\cite{Dubovsky:2004sg} which is minimal in the sense that in does not
contain new light propagating degrees of freedom as compared to the
Einstein gravity. In this model the graviton acquires a mass due to
the space-time dependent condensates of the four ``Goldstone'' scalars
$\phi^0(x)$, $\phi^i(x)$. The action of the model reads
\cite{Dubovsky:2004ud}
\begin{equation}
\mathcal{S} = \int \textrm{d}^4 x \sqrt{- g} \left[ -
M^2_{Pl} \mathcal{R} + \Lambda^4 \mathcal{F} \left( Z^{ij}
\right) + {\mathcal L}_{\rm matter}\right],
\label{eq:MGaction}
\end{equation}
where the first term is the standard Einstein-Hilbert action,
${\mathcal L}_{\rm matter}$ stands for the minimally coupled ordinary
matter and $\mathcal{F}(Z^{ij})$ is a function of the derivatives of the
four scalar fields $\phi^0(x)$, $\phi^i(x)$ which depends on a single
argument $Z^{ij}$ constructed as follows:
\begin{eqnarray} \label{XVW}
\nonumber
Z^{ij} &=& X^{\gamma}W^{ij},\\
X &=& \Lambda^{-4} g^{\mu\nu} \partial_\mu \phi^0 \partial_\nu \phi^0, \nonumber \\
V^i &=& \Lambda^{-4} g^{\mu\nu} \partial_\mu \phi^0 \partial_\nu \phi^i, \nonumber \\
W^{ij} &=& \Lambda^{-4} g^{\mu\nu} \partial_\mu \phi^i \partial_\nu \phi^j
- \frac{V^i V^j}{X} \, .
\end{eqnarray}
The constant $\gamma$ is a free parameter. The model possesses
meaningful cosmological solutions for $\gamma \geq 1/3$ and $\gamma<0$
\cite{Dubovsky:2005dw}.

The flat space vacuum solution is \footnote{To be precise, the vacuum
solution reads $\phi^0=at$, $\phi^i=bx^i$, where $a$ and $b$ are two
constants whose value is set by the requirement that the
energy-momentum of the scalar fields in the Minkowski space
vanishes. We assume in what follows that the function $\mathcal{F}$
is such that $a=b=\Lambda^2$.}
\[
\phi_0 = \Lambda^2 t, \qquad \phi^i = \Lambda^2 x^i.
\]
The ground state is translationally invariant due to the derivative
nature of the Goldstone coupling, but the Lorentz symmetry is
spontaneously broken. The particular dependence of the action
(\ref{eq:MGaction}) on the derivatives of the Goldstone fields through
a single argument $Z^{ij}$ ensures the non-pathological behavior of
the perturbations about the vacuum solution \cite{Dubovsky:2004sg},
namely, the absence of ghosts and rapid instabilities. The tensor
metric perturbations are, in general, massive with the mass determined
by the first and the second derivatives of the function~$\mathcal{F}$.
The low-energy spectrum consists of the two propagating tensor
modes. The auxiliary scalars do not appear in the spectrum.

An interesting and peculiar feature of this model is that the
gravitational interaction between static sources is described by the
standard Newton's law despite the non-zero mass of the
graviton\footnote{This feature is obviously a consequence of the
breakdown of the Lorentz invariance.}. Due to this feature the model passes
the terrestrial and solar system tests even for graviton masses as
large as $(10^{15}{\rm cm})^{-1}$. Moreover, it admits the standard
Friedmann-Robertson-Walker cosmological solutions, the only trace of
the Goldstone scalars being two contributions to the energy density
which behave as a cosmological constant and as matter with the
equation of state $w=-(3\gamma)^{-1}$. The massive gravitons may be
created in cosmologically significant amount and may play a role of
the dark matter \cite{Dubovsky:2004ud}.

Given that the model passes the most obvious constraints one may
wonder if it reproduces correctly more subtle parts of modern
cosmology, in particular, the theory of structure formation. This is
the question which we address in this paper. The answer is not obvious
{\em a priori} since the vacuum in this model contains the condensates
of the Goldstone fields whose perturbations mix with the matter
density perturbations.

We find that cosmological perturbations in the model
(\ref{eq:MGaction}) consist of two parts. The first part behaves
identically to the perturbations in GR and is, therefore, compatible
with observations. The second, ``anomalous'' part is proportional to
an unknown function $\vartheta(x^i)$ of the space coordinates which
arises as an integration constant.

The appearance of this function reflects the existence of the
non-dynamical mode with the dispersion relation $\omega^2=0$, much in
common with the ghost condensate model \cite{Arkani-Hamed:2003uy}. In
the model (\ref{eq:MGaction}) the value of $\vartheta(x^i)$ is
determined by the initial conditions. Note, however, that the action
(\ref{eq:MGaction}) is no more than the low-energy effective
action. One should expect the higher-derivative corrections to
eq.~(\ref{eq:MGaction}) to be present. In general, these corrections
make $\vartheta(x^i)$ a slowly changing dynamical variable. In this
case the initial value of $\vartheta(x^i)$ may be determined by its
evolution at the inflationary epoch.

As we show below, the growth of the ``anomalous'' perturbations
depends on the value of $\gamma$. For $-1<\gamma<0$ they grow slower
than the standard ones, so that the latter dominate. At $\gamma=1$ the
anomalous contributions to perturbations cancel out. Thus, at least in
these two cases the perturbations behave in a standard way and the
model (\ref{eq:MGaction}) is consistent with the structure formation
in the Universe. This is the main result of the paper.

For other values of $\gamma$ (in particular, for the cosmologically
interesting case $1/3<\gamma<1$ which corresponds to the equation of
state of the condensate $-1<w<-1/3$) the anomalous perturbations {\em
grow} at the radiation-dominated epoch, and may or may not grow {\em
faster} than the standard perturbations at the matter-dominated stage,
depending on the particular value of $\gamma$ (see
Sect.~\ref{sec:cosm-pert} for details). Whether this leads to a
contradiction with observations depends on the function
$\vartheta(x^i)$ at the beginning of the radiation-dominated era. To
address this question one needs to consider the higher-derivative
corrections to the action (\ref{eq:MGaction}). This will be done
elsewhere.

The paper is organized as follows. In Sect.~\ref{sec:frw-solution} we
review the cosmological solutions in massive gravity. In
Sect.~\ref{sec:cosm-pert} we calculate the growth of perturbations in
the FRW background. Sect.~\ref{sec:discussion} contains the summary of
the results and their discussion. The details of the calculations are
given in the Appendix.

\section{FRW solution}
\label{sec:frw-solution}

In the absence of matter the model described by the action
(\ref{eq:MGaction}) admits the standard FRW solutions
\cite{Dubovsky:2005dw}. The spatially-flat homogeneous and isotropic
ansatz \footnote{This ansatz cannot be directly generalized to the cases
  of open and closed Universe \cite{Dubovsky:2004sg}. It is not known
  whether the model admits cosmological solutions of these types. }
for the metric and Goldstone fields reads
\begin{eqnarray} \label{FRW}
\nonumber
ds^2 &=& a^2 (\eta) \left( d\eta^2 - \delta_{ij} dx^i dx^j\right),\\
\phi^0 &=& \Lambda^2 \phi \left( \eta \right), \qquad \phi^i = \Lambda^2 x^i,
\label{eq:ansatz}
\end{eqnarray}
which implies
\begin{equation*}
Z = \phi'^{2\gamma}/a^{2\gamma+2},
\label{eq:Z=}
\end{equation*}
where $Z = - \delta_{ij} Z^{ij} / 3$ and where prime denotes the
derivative with respect to the conformal time $\eta$. Assuming the
ordinary matter is homogeneous and isotropic, the equations of motion
(the Friedmann equation and the field equation for $\phi^0$) take the
form
\begin{eqnarray}
\mathcal{H}^2 &=& {a^2 \over 3 M_{\rm Pl}^2} 
\left( \rho_m + \rho_\phi + \rho_\Lambda \right),
\label{eq:Friedmann} \\
0 &=& \partial_0 \left( a^{3-1/\gamma} Z^{1-1/2\gamma} \mathcal{F}_Z \right) ,
\label{eq:field}
\end{eqnarray}
where $\mathcal{H}=a'/a$, so that $\mathcal{H}/a$ is the Hubble
constant, and $3 \mathcal{F}_Z = \delta^{ij}
d\mathcal{F}/dZ^{ij}$. The matter energy density $\rho_m$ has the
standard form, while the two contributions of the Goldstone fields to
the energy density read
\begin{eqnarray*}
\rho_\Lambda = - \Lambda^4 \mathcal{F} / 2, && 
\rho_\phi = - 3 \gamma \Lambda^4 Z \mathcal{F}_Z .
\end{eqnarray*}
The first of these terms behaves like a cosmological constant, while
the second contribution corresponds to matter with the equation of
state $w = - 1/( 3 \gamma)$. Thus, the Friedmann equation
(\ref{eq:Friedmann}) reduces to the standard one, the only trace of
the Goldstone field being the energy densities $\rho_\phi$ and
$\rho_\Lambda$.

For a given function ${\cal F}$  eq.~(\ref{eq:field}) determines the
dependence of the variable $Z$ on the scale factor.  For $\gamma >
1/3$ or $\gamma < 0$ this equation implies
\begin{eqnarray*}
Z^{1-1/2\gamma} \mathcal{F}_Z \rightarrow 0 &\textrm{as}& a
\rightarrow \infty .
\end{eqnarray*}
The case of interest (which occurs generically for an algebraic
function ${\cal F}$ \cite{Dubovsky:2005dw}) is when in this limit $Z \to Z_0={\rm const.}$
such that ${\cal F}_Z(Z_0)=0$. In this case the graviton mass
remains finite at $a\to\infty$ (see Sect.~\ref{sec:mass-parameters}
for details). 

In a special case $\gamma = 1 / 3$ the field equation (\ref{eq:field})
implies $Z = \textrm{const}$.  Then $\mathcal{F}_Z$ is not driven to
zero by the cosmological evolution and $\rho_\phi$ behaves as the
cosmological constant. The value of the total cosmological constant
$\rho_\phi+\rho_\Lambda$ is determined by the initial conditions.

In what follows we neglect where possible the deviations from the
point $Z = Z_0$ (note that in this point $\rho_\phi =0$). We also
assume that the value of $\rho_\Lambda$ is of the order of the
present-day cosmological constant, and thus its contribution to the
Friedmann equation at the epoch of structure formation is negligible.

\section{Cosmological perturbations}
\label{sec:cosm-pert}

As in the standard analysis of the cosmological perturbations
\cite{Mukhanov:1990me}, it is convenient to separate space and time
components. The metric perturbations can be parameterized in the
following way,
\begin{eqnarray}
\nonumber
\delta g_{00} &=& 2 a^2 \varphi,\\
\delta g_{i0} &=& a^2 (v_i + \partial_i B), \label{metric-pert}\\
\nonumber
\delta g_{ij} &=& a^2 (2 \psi \delta_{ij} 
- \partial_i F_j - \partial_j F_i - 2 \partial_i \partial_j E +
h_{ij}),
\end{eqnarray}
where the vector perturbations $v_i$ and $F_i$ are transverse, while
the tensor perturbation $h_{ij}$ is transverse and traceless.  A
similar parameterization can be used for the perturbations of the
Goldstone fields,
\begin{eqnarray}
\nonumber
\delta\phi^0&=&\Lambda^2 \xi^0,\\
\delta \phi^i &=& \Lambda^2( \xi_i + \partial_i \xi), \label{Gold-pert}
\end{eqnarray}
where $\xi_i$ is transverse.  Finally, the perturbations of the
ordinary matter are parameterized in the following way,
\begin{eqnarray}
\nonumber
\delta \mathcal{T}_{\mu\nu}^{m} &=& \left( \delta \rho_m + \delta p_m
\right) u_{\mu} u_{\nu} - g_{\mu\nu} \delta
p_m - p_m \delta g_{\mu\nu} \\ && + \left( \rho_m + p_m \right) \left( u_{\nu}
\delta u_{\mu} + u_{\mu} \delta u_{\nu} \right) , \label{energy-tensor-pert}
\end{eqnarray}
where $\delta \rho_m$ and $\delta p_m$ are related by the matter
equation of state and the perturbations of the velocity $\delta u_\mu$
are expressed in terms of the scalar $\zeta$ and the transverse vector
$\zeta_i$ as follows, 
\begin{eqnarray*}
\nonumber
\delta u_i &=& a (\zeta_i + \partial_i \zeta),\\
\delta u_0 &=& a\varphi.
\end{eqnarray*}
Therefore, in total there is one tensor perturbation $h_{ij}$
consisting of two components, four vectors $v_i$, $F_i$, $\xi_i$ and
$\zeta_i$, consisting of two components each, and 9 scalars $\varphi$,
$B$, $\psi$, $E$, $\xi^0$, $\xi$, $\zeta$, $\delta p_m$ and $\delta
\rho_m$. One vector and two scalar perturbations are gauge degrees of
freedom; they can be eliminated by imposing a gauge condition. As a
consequence, there is only three gauge-invariant vector fields
\begin{eqnarray} \label{gauge-vector}
\varpi_i = v_i + F_i', && \sigma_i = \xi_i - F_i ,
\end{eqnarray}
and $\zeta_i$, and seven scalar gauge-invariant fields
\begin{eqnarray}
\Phi &=& \varphi - a^{-1} \left[ a \left( E^\prime + B \right) \right]', \nonumber \\
\Psi &=& \psi + \mathcal{H} \left( E^\prime + B \right) , \nonumber \\
\Xi &=& \xi - E, \nonumber \\
\mathcal{B} &=& B + E^\prime - \xi^{0} / \phi^{\prime} , \nonumber \\
\delta_\rho &=& \left[ \delta\rho_m 
- \rho_m^\prime \left( E^\prime + B \right) \right] / \rho_m, \nonumber \\
\delta_p &=& \left[ \delta p_m - p_m^\prime 
\left( E^\prime + B \right) \right] / p_m, \nonumber \\
\delta_\zeta &=& \zeta - \left( E^\prime + B \right) . \label{gauge-scalars}
\end{eqnarray}
The tensor perturbation $h_{ij}$ is also gauge invariant.

\subsection{The tensor perturbations}
\label{sec:tensor-perturbations}

The equation for the tensor perturbation is
\begin{eqnarray} \label{021}
h_{ij}'' + 2 \mathcal{H} h_{ij}' 
- \partial_i^2 h_{ij} + a^2 m_2^2 h_{ij} &=& 0,
\end{eqnarray}
where $m_2$ is the graviton mass which has the scale $m_2
\propto \Lambda^2/M_{\rm Pl}$ and whose precise expression in terms of
the function $\mathcal{F}(Z^{ij})$ and its derivatives is given in the
Appendix \ref{sec:mass-parameters}.

This equation is identical to the equation for a free massive scalar
field in the FRW background. If the mass of the graviton is larger
than the Hubble constant, $m_2 \gg \mathcal{H}/a$, which we assume to
be the case in what follows, eq.~(\ref{021}) describes massive
gravitational waves with the amplitude scaling like $h_{ij} \propto
a^{-1}$ and $h_{ij} \propto a^{-3/2}$ in the relativistic and
non-relativistic limits, respectively.

\subsection{The vector perturbations}
\label{sec:vector-perturbations}

In the longitudinal gauge the three equations describing vector
perturbations are
\begin{eqnarray} \label{025}
&& \partial_j^2 \varpi_i - 2 a^2 \rho_m M_{pl}^{-2} ( 1 + w) 
\zeta_i = 0,\\ \label{025a}
&& \varpi_i' + 2 \mathcal{H} \varpi_i - a^2 m_2^2 \sigma_i = 0,
\\ \label{025b}
&& m_2^2 \partial_j^2 \sigma_{i} = 0,
\end{eqnarray}
where $w$ is the parameter entering the equation of state of the
ordinary matter, $p_m = w\rho_m$.

The first of these equations allows to express $\zeta^i$ in terms of
$\varpi_i$, while the third equation gives $\sigma_i = 0$. Therefore,
the only non-trivial equation is eq.~(\ref{025a}). It differs from the
conventional one by the term proportional to the graviton mass $m_2^2$
which cancels at $\sigma_i = 0$. Thus, this equation is the conventional
one and describes a field with the amplitude decreasing as $\varpi_i
\propto a^{-2}$.

\subsection{The scalar perturbations}
\label{sec:scalar-perturbations-1}

The perturbation $\delta_p$ can be expressed in terms of
$\delta_\rho$ by means of the matter equation of state. In the case
of the adiabatic perturbations one has
\[
\delta_p = \dfrac{c_s^2}{w} \delta_\rho ,
\]
where $c_s$ is the sound velocity ($c_s^2=w$ for the ideal fluid).  
The behavior of the remaining six perturbations is governed
by the following equations,
\begin{eqnarray}
0 &=& \Phi - \Psi + a^2 m_2^2 \Xi \, , \label{029} \\
0 &=& 2 \left( \Psi^\prime + \mathcal{H} \Phi \right) - a^2 \frac{\rho_m}{M_{Pl}^2} \left( 1 + w_\omega \right) \delta_\zeta , \label{030} \\
0 &=& - 2 \partial_j^2 \Psi + 6 \mathcal{H} \left( \mathcal{H} \Phi + \Psi^\prime \right) + a^2 m_4^2 \left( \partial_j^2 \Xi + 3 \Psi \right) \nonumber \\
& & + a^2 \frac{\rho_m}{M_{Pl}^2} \delta_\rho - a^2 m_0^2 \left[ \frac{\phi^{\prime\prime}}{\phi^{\prime}} \mathcal{B} + \mathcal{B}^{\prime} + \Phi \right] , \label{031} \\
0 &=& - 2 \Psi^{\prime\prime} - 2 \Phi \left( \mathcal{H}^2 + 2 \mathcal{H}^\prime \right) + \partial_j^2 \left( \Psi - \Phi \right) \nonumber \\
& & - 2 \mathcal{H} \left( 2 \Psi + \Phi \right)^\prime + a^2 \left[ \frac{p_m}{M_{Pl}^2} \delta_p - m_3^2 \partial_j^2 \Xi \right. \nonumber \\
& & \left. + m_4^2 \left( \Phi + \frac{\phi^{\prime\prime}}{\phi^{\prime}} \mathcal{B} + \mathcal{B}^{\prime} - \Psi / \gamma \right) \right] , \label{032}
\end{eqnarray}
\begin{eqnarray}
0 &=& \partial_0 \left[ \frac{a^4 m_4^2}{\phi^{\prime}} \left( 3 \gamma \left( \Phi + \mathcal{B}^\prime + \frac{\phi^{\prime\prime}}{\phi^{\prime}} \mathcal{B} \right) \right. \right. \nonumber \\
& & \left. \left. - \left( 3 \Psi + \partial_i^2 \Xi \right) \right) \right] , \label{033} \\
0 &=& m_4^2 \left( \Psi / \gamma - \frac{\phi^{\prime\prime}}{\phi^{\prime}} \mathcal{B} - \mathcal{B}^\prime - \Phi \right) \nonumber \\
& & + \left( m_3^2 - m_2^2 \right) \partial_j^2 \Xi . \label{034}
\end{eqnarray}
Here $m_i$ are the graviton mass parameters defined in the Appendix
\ref{sec:mass-parameters}. We assume that they are of the
order of $\Lambda^2/M_{\rm Pl}$.

This system of equations can be solved as follows. At $m_4^2\neq 0$
eq.~(\ref{034}) can be used to express $\phi^{\prime\prime} /
\phi^{\prime} \mathcal{B} + \mathcal{B}^\prime + \Phi$. Then
eq.~(\ref{033}) becomes a closed equation for $\Xi$
\begin{eqnarray*}
0 &=& \partial_i^2 \left[ \Xi' + \left( 3 - \frac{1}{\gamma} \right) \mathcal{H} \Xi \right] .
\end{eqnarray*}
The solution of this equation which does not grow at spatial infinity 
reads
\begin{eqnarray} \label{eq:theta=xi}
m_2^2 \, \Xi = - \vartheta \left( x^i \right) \, a^{1/\gamma - 3} ,
\end{eqnarray}
where $\vartheta \left( x^i \right)$ is an integration constant
depending only on spatial coordinates.
Then eqs.~(\ref{030}), (\ref{031}) and (\ref{034}) can be used
to express $\delta_\zeta$, $\delta_\rho$ and $\mathcal{B}$ in terms of
$\Phi$ and $\Psi$ while eq.~(\ref{029}) reads
\begin{eqnarray}
\Phi - \Psi = \vartheta \left( x^i \right) \, a^{1/\gamma - 1}.
\label{eq:036}
\end{eqnarray}
With the account of all these relations, the remaining
eq.~(\ref{032}) becomes a closed inhomogeneous equation for $\Psi$,
\begin{widetext}
\begin{eqnarray}
0 &=& \partial_a^2 \Psi + \frac{1}{a} \left( 4 + 3 c_s^2 
+ \frac{\mathcal{H}^\prime}{\mathcal{H}^2} \right) \partial_a \Psi 
+ \frac{1}{a^2} \left[ \left( 1 + 3 c_s^2 \right) 
+ 2 \frac{\mathcal{H}^\prime}{\mathcal{H}^2} 
- \frac{c_s^2 \partial_i^2}{\mathcal{H}^2} \right] \Psi \nonumber \\
& & - \left[ \frac{\gamma c_s^2 \partial_i^2}{\mathcal{H}^2} 
- \left( 3 c_s^2 + \frac{1}{\gamma} 
+ 2 \frac{\mathcal{H}^\prime}{\mathcal{H}^2} \right) \right] \vartheta 
\, a^{1/\gamma - 3} \label{037} .
\end{eqnarray}
\end{widetext}
Once the solution to this equation is found, the other variables are
determined by eqs.~(\ref{eq:036}), (\ref{030}),
(\ref{031}) and (\ref{034}). In particular, one finds
\begin{eqnarray}
\delta_\rho &=& \frac{2 M_{pl}^{2}}{\rho_m} \left(
\gamma \partial_i^2 - 3 \mathcal{H}^2 \right) 
a^{1/\gamma-3} \vartheta \label{eq:drho/rho=}
\\ &&
- \frac{2 M_{pl}^{2}}{a^2 \rho_m} \left[ 3
\mathcal{H}^2 \left( 1 + a \frac{\partial}{\partial a} \right) -
\partial_i^2 \right] \Psi, 
\nonumber
\end{eqnarray}
where $\Psi$ is a solution to eq.~(\ref{037}).

The conventional cosmological perturbations are recovered by setting
the graviton masses to zero, $m_i^2=0$. In this case eq.~(\ref{029})
gives $\Phi-\Psi=0$ which implies $\vartheta(x^i)=0$
(cf. eqs.~(\ref{eq:theta=xi}) and (\ref{eq:036})). Then both
eq.~(\ref{037}) and eq.~(\ref{eq:drho/rho=}) reduce to the standard
equations describing cosmological perturbations in the Einstein
theory. Note that the value of $\vartheta$ is determined essentially by
the initial conditions. Setting $\vartheta=0$ would eliminate the
$\vartheta$-dependent terms in eqs.~(\ref{037}) and (\ref{eq:drho/rho=})
and bring these equations to the conventional form even in the case
$m_2^2\neq 0$.

In the case of matter perturbations in a matter-dominated Universe
eq.~(\ref{037}) reduces to the following equation,
\begin{eqnarray*}
\frac{\partial^2 \Psi}{\partial a^2} + \frac{7}{2 a} 
\frac{\partial \Psi}{\partial a} 
+ \left( \frac{1}{\gamma} - 1 \right) a^{1/\gamma - 3} \vartheta= 0,
\label{eq:matter-domination}
\end{eqnarray*}
which differs from the standard case by the presence of the
inhomogeneous term proportional to $\vartheta$. The solution to this
equation reads
\begin{eqnarray*} \label{043}
\Psi = - \frac{2 \gamma}{2 + 3 \gamma} a^{1/\gamma - 1} \vartheta(x^i) 
+ a^{-5/2}  c_1(x^i)
+ c_2(x^i),
\end{eqnarray*}
where $c_i(x^i)$ are the integration constants. Substituting this
solution into eq.~(\ref{eq:drho/rho=}) one finds the density contrast 
\begin{eqnarray}
\nonumber
\delta_\rho
&=& \left( {2 M_{pl}^{2} a \over \rho_0} 
\partial_i^2 + 3 \right) \dfrac{c_1(x^i)}{a^{5/2}}
+ 2 \left( {a M_{pl}^{2} \over \rho_0} \partial_i^2 -1\right) c_2(x^i)
\\ && \label{eq:drho/rho-matter} + {6 \gamma \over 2 + 3 \gamma} 
a^{1/\gamma - 1} \left( {a \gamma M_{pl}^{2} \over \rho_0}
\partial_i^2 - 1 \right) \vartheta (x^i).
\end{eqnarray}
where $\rho_0$ is the energy density of matter at present. The first
two terms in this equation are precisely the ones which appear in the
standard Einstein theory, the second term describing the linear growth
of the perturbations, $\delta_\rho \propto a$.  The difference with
the conventional case consists in the third term on the r.h.s of
eq.~(\ref{eq:drho/rho-matter}). The perturbations corresponding to
this term grow proportionally to $a^{1/\gamma}$. For $\gamma > 1$ or
$\gamma < 0$ these ``anomalous'' perturbations grow slower than the
standard ones.

In the radiation epoch the situation is similar. For a relativistic
fluid one has $w_m = 1 / 3$, so that eq.~(\ref{037}) reduces to the
following one,
\begin{eqnarray}
0 &=& \frac{\partial^2 \Psi}{\partial a^2} + \frac{4}{a}
\frac{\partial \Psi}{\partial a} - \frac{M_{pl}^2
\partial_i^2}{\rho_r} \Psi \nonumber \\
& & + \left( \frac{1}{\gamma} - 1 - \frac{a^2 \gamma M_{pl}^2
\partial_i^2}{\rho_r} \right) a^{1/\gamma-3}\vartheta,
\label{eq:psi_radiation}
\end{eqnarray}
where $\rho_r$ is the energy density of radiation at present.  For the
generic value of $\gamma$ the solution to this equation is
cumbersome. For simplicity let us concentrate on the modes which are
much smaller than the Hubble scale, $k^2 \gg \mathcal{H}^2$. The
density contrast calculated according to eq.~(\ref{eq:drho/rho=}) has
the standard oscillating piece and the extra part proportional to
$\vartheta$, 
\begin{eqnarray}
\nonumber
\delta_\rho &\sim& c_1(x^i) \sin y + c_2(x^i) \cos y 
+ 2 \gamma \left( \dfrac{\rho_r}{ k^2 M_{pl}^2}
\right)^{(1/\gamma-1)/2} 
\\ \label{051} 
& & \times
\left[ - y^{1+1/\gamma}+ \int_0^y \textrm{d}x \, 
x^{1+1/\gamma} \sin (y-x) \right] 
 \vartheta ,\
\end{eqnarray}
where $y = \eta k / \sqrt{3}$ is proportional to the scale factor,
while $c_i(x^i)$ are two integration constants. As one
may see from this expression, for $-1\leq \gamma<0$ the
$\vartheta$-dependent contribution to the density contrast decays with
the scale factor so that only the standard contribution remains. Thus,
in this range of $\gamma$ the perturbations behave just as predicted
by general relativity in both matter and radiation-dominated epochs.

Another case of interest is $\gamma=1$. This case is special because
at $\gamma=1$ the $a$-dependence of the last term in
eq.~(\ref{eq:psi_radiation}) disappears. In fact, one may show that in
this case the dependence on $\vartheta$ cancels out in the density
contrast, so that only the standard part of perturbations
remains. 

At other values of $\gamma$ the $\vartheta$-dependent contributions to
perturbations grow in the radiation-dominated Universe. 

\section{Discussion}
\label{sec:discussion}

To summarize, in the model of massive gravity described by the action
(\ref{eq:MGaction}) the cosmological perturbations contain two parts,
the ``normal'' and the ``anomalous'' one. The first, normal part has
the behavior identical to that found in the conventional general
relativity. It is therefore in agreement with observations to the same
extent as the latter. In particular, the ``normal'' part of the
perturbations can describe successfully at least the linear stage of
the structure formation. In fact, the predictions of the general
relativity and the model (\ref{eq:MGaction}) with only the
``normal'' perturbations present completely coincide in the linear
regime, so that the two models are indistinguishable in this respect.

The second, ``anomalous'' part of perturbations is specific to the
model of massive gravity with the action (\ref{eq:MGaction}). These
perturbations originate from the condensates of the scalar fields
present in the model. This contribution to perturbations depends
linearly on the unknown function of space coordinates $\vartheta(x^i)$
which enters as an integration constant. The value of this function
cannot be determined within the model (\ref{eq:MGaction}) and has to
be specified as an initial conditions.

The behavior of the ``anomalous'' contributions to perturbations at
different stages of the evolution of the Universe depends on the value
of $\gamma$. At the matter-dominated stage the anomalous perturbations
grow not faster than the standard ones for $\gamma\geq 1$ and
$\gamma<0$. In the radiation-dominated epoch this occurs at $-1\leq
\gamma<0$ and $\gamma=1$. Thus, at $-1\leq \gamma<0$ and $\gamma=1$
the normal perturbations dominate at both radiation and
matter-dominated stages. 

The appearance of the time-independent arbitrary function is not
surprising. The same function $\vartheta(x^i)$ also enters the
expression for the gravitational potential of an isolated massive body
\cite{Dubovsky:2004ud}. Its origin may be traced back to the existence of the
scalar mode with the dispersion relation $\omega^2=0$
\cite{Dubovsky:2004sg}. This mode is not dynamical in the model
(\ref{eq:MGaction}). However, the action (\ref{eq:MGaction}) is the
low-energy effective action, so one should expect the corrections
containing higher-derivative terms to be present. In general, these
corrections make $\vartheta$ a dynamical variable with the dispersion
relation $\omega^2=\alpha p^4$, where $\alpha$ is a small
coefficient. Therefore, $\vartheta$ becomes a slowly varying function
of time. The slow evolution may drive $\vartheta$ to a particular
value at the inflationary epoch and thus prepare the initial
conditions for the radiation-dominated stage. If this initial value of
$\vartheta$ is small, then the growth of the ``anomalous'' part of
perturbations may become irrelevant and corresponding values of
$\gamma$ phenomenologically acceptable. This question will be
considered elsewhere.

\begin{acknowledgments}
The authors are grateful to S.~Dubovsky for valuable discussions and
reading of the manuscript. The work of M.B. is supported by the
Belgian FRIA, Fond pour la Formation \`a la Recherche dans l'Industrie
et dans l'Agronomie. The work of P.T. is supported by IISN, Belgian
Science Policy (under contract IAP V/27).
\end{acknowledgments}

\appendix

\section*{Appendix}
\label{sec:appendix}

In these Appendix we provide some intermediate formulae skipped in
the main text. In Appendix \ref{sec:background} we calculate the field
equations for the standard FRW solution (\ref{FRW}). In Appendix
\ref{sec:mass-parameters} we introduce the mass parameters and the
relations between them. In Appendix \label{sec:linearized-equations}
we calculate the gauge-invarant linearized field equations.

\section{The background}
\label{sec:background}

For $X$, $V^i$ and $W^{ij}$ given by eq.~(\ref{XVW}), the following
action describes a massive gravitational field
\begin{eqnarray*} \label{A01}
\mathcal{S} &=& \int \textrm{d}^4 x \sqrt{- g} 
\left[ - M^2_{Pl} \mathcal{R} + \Lambda^4 \mathcal{F} 
\left( X , V^i , W^{ij} , \ldots \right) \right] ,
\end{eqnarray*}
characterized by the ten Einstein equations $\mathcal{G}_{\mu\nu} =
M^{-2}_{pl} \left( \mathcal{T}_{\mu\nu}^m +
\mathcal{T}_{\mu\nu}^\phi \right)$ and four Goldstone equations
\begin{eqnarray*} \label{A02}
0 &=& \partial_\beta \left\lbrace \sqrt{- g} g^{\alpha\beta} \left[
\frac{\partial \mathcal{F}}{\partial X} \delta^0_\mu \partial_\alpha
\phi^{0} + \frac{\partial \mathcal{F}}{\partial W^{ij}} \left( \delta^i_\mu
\partial_\alpha \phi^{j} \right. \right. \right. \nonumber \\
& & \left. - \frac{V^{j}}{X} \partial_\alpha \left(
\phi^{0} \delta^i_\mu + \phi^{i} \delta^0_\mu \right)
+ \delta^0_\mu \frac{V^{i} V^{j}}{X^2} \partial_\alpha \phi^{0} \right) \nonumber \\
& & \left. \left. + \frac{1}{2} \frac{\partial \mathcal{F}}{\partial V^{i}}
\partial_\alpha \left( \phi^{0} \delta^i_\mu + \phi^{i} \delta^0_\mu
\right) \right] \right\rbrace .
\end{eqnarray*}
Here $\mathcal{T}_{\mu\nu}^m$ is the energy-momentum tensor of the
ordinary matter which has the standard form (\ref{energy-tensor-pert})
while the energy-momentum tensor of the Goldstone field is given by
\begin{eqnarray*} \label{A04}
\mathcal{T}_{\mu\nu}^\phi &=& \Lambda^{4} \left\{ - \frac{1}{2}
g_{\mu\nu} \mathcal{F} + \frac{1}{2}
\frac{\partial \mathcal{F}}{\partial V^{i}} \left( \partial_{\mu}
\phi^{i} \partial_{\nu} \phi^{0} + \partial_{\mu} \phi^{0}
\partial_{\nu} \phi^{i} \right) \right. \nonumber \\ & & \left. +
\left[ \partial_{\mu}
\phi^{i} \partial_{\nu} \phi^{j} - \frac{V^j}{X} \left(
\partial_{\mu} \phi^{i} \partial_{\nu} \phi^{0} + \partial_{\mu}
\phi^0 \partial_{\nu} \phi^i \right) \right. \right. \nonumber \\ & &
\left. \left. + \frac{V^i V^j}{X^2} \partial_{\mu} \phi^0
\partial_{\nu} \phi^0 \right] \frac{\partial \mathcal{F}}{\partial W^{ij}} + \frac{\partial \mathcal{F}}{\partial X}
\partial_{\mu} \phi^{0} \partial_{\nu} \phi^{0} \right\} .
\end{eqnarray*}
For the cosmological solutions given by eq.~(\ref{eq:ansatz}), $X =
a^{-2} \phi^{\prime2}$, $V^i = 0$ and $W^{ij} = - a^{-2}
\delta^{ij}$. The 14 field equations reduce to the following three
relations (prime stands for the derivative with respect to $\eta$),
\begin{eqnarray*}
3 \mathcal{H}^2 &=& \dfrac{a^2}{M_{pl}^2} \left( \rho_m + \rho_\phi + \rho_\Lambda \right) , \nonumber
\\ 2 \mathcal{H}^\prime + \mathcal{H}^2 &=& - \dfrac{a^2}{M_{pl}^2} \left( p_m + p_\phi + p_\Lambda
\right) , \label{A06} \\ 0 &=& \partial_0 \left( a^3 \mathcal{F}_X
X^{1/2} \right) , \nonumber
\end{eqnarray*}
which can be solved for any given function $\mathcal{F} \left( a , X
\right)$. The following notations were used
\begin{eqnarray*} \label{A07}
\begin{array}{lcl}
\rho_\phi = \Lambda^{4} X \mathcal{F}_X , & & p_\phi =
\Lambda^{4} W \mathcal{F}_W , \\
\rho_\Lambda = - \Lambda^{4} \mathcal{F} / 2 , & & p_\Lambda = \Lambda^{4} \mathcal{F} / 2 .
\end{array}
\end{eqnarray*}
For the model (\ref{eq:MGaction}) characterized by a function ${\cal F}$ depending only on $Z^{ij}$
 one has  $p_\phi = - \left( 3 \gamma \right)^{-1} \rho_\phi$.

\section{The mass parameters}
\label{sec:mass-parameters}

There are five mass parameters $m_i^2$, $i = 0 \ldots 4$ defined by
the following relations
\begin{eqnarray*}
m_0^2 &=& \frac{\Lambda^{4}}{M^2_{Pl}} \left( X \mathcal{F}_{X} + 2 X^2 \mathcal{F}_{XX} \right) ,
\nonumber \\
m_1^2 &=& \frac{2 \Lambda^{4}}{M^2_{Pl}} \left( - X \mathcal{F}_{X} - W \mathcal{F}_{W} + \frac{1}{2} X W \mathcal{F}_{VV} \right) ,
\nonumber \\
m_2^2 &=& \frac{2 \Lambda^{4}}{M^2_{Pl}} \left( W \mathcal{F}_{W} - 2 W^2 \mathcal{F}_{WW2} \right) ,
\nonumber \\
m_3^2 &=& \frac{\Lambda^{4}}{M^2_{Pl}} \left( W \mathcal{F}_{W} + 2 W^2 \mathcal{F}_{WW1} \right) ,
\nonumber \\
m_4^2 &=& - \frac{\Lambda^{4}}{M^2_{Pl}} \left( X \mathcal{F}_{X} + 2 X W \mathcal{F}_{XW} \right) , \label{A20}
\end{eqnarray*}
where $W = -1 / 3 \delta_{ij} W^{ij}$ and where the first and second
derivatives of the function $\mathcal{F} \left( X , V^i , W^{ij} ,
\ldots \right)$ are denoted as follows
\begin{eqnarray*}
\frac{\partial \mathcal{F}}{\partial X} &\equiv& \mathcal{F}_{X}, \\
\frac{\partial^2 \mathcal{F}}{\partial X^2} &\equiv& \mathcal{F}_{XX} , \\
\frac{\partial^2 \mathcal{F}}{\partial V^{i}V^{j}} &\equiv& \mathcal{F}_{VV} \delta_{ij}, \\
\frac{\partial \mathcal{F}}{\partial W^{ij}} &\equiv& \mathcal{F}_{W} \delta_{ij} , \\
\frac{\partial^2 \mathcal{F}}{\partial X W^{ij}} &\equiv& \mathcal{F}_{XW} \delta_{ij}, \\
\frac{\partial^2 \mathcal{F}}{\partial W^{ij} W^{kl}} &\equiv& \mathcal{F}_{WW1} \delta_{ij} \delta_{kl} + \mathcal{F}_{WW2} \left( \delta_{ik} \delta_{jl} + \delta_{il} \delta_{jk} \right) .
\end{eqnarray*}
It is possible to relate the masses $m_0^2$ and $m_4^2$ by using the Goldstone equation for the background. This last gives
\begin{eqnarray*}
m_0^2 \left( \frac{\phi^{\prime\prime}}{\phi^{\prime}} - \mathcal{H} \right) = 3 \mathcal{H} m_4^2 .
\end{eqnarray*}

For the model characterized by the function $\mathcal{F} = \mathcal{F}
\left( Z^{ij} \right)$, it is straightforward to show that
\begin{eqnarray*}
m_0^2 &=& \dfrac{\Lambda^{4}}{\mathcal{M}^2_{pl}} \gamma 
[ 3 ( 1 - 2 \gamma ) Z \mathcal{F}_{Z} \\
&& + 6 \gamma Z^2 
\left( 3 \mathcal{F}_{ZZ1} + 2 \mathcal{F}_{ZZ2} \right) ] , \nonumber \\
m_1^2 &=& \frac{2 \Lambda^{4}}{M^2_{pl}} \left( 3 \gamma - 1 \right) 
Z \mathcal{F}_{Z} , \nonumber
\\
m_2^2 &=& \dfrac{2 \Lambda^{4}}{\mathcal{M}^2_{pl}} 
\left( Z \mathcal{F}_{Z} - 2 Z^2 \mathcal{F}_{ZZ2} \right) , \nonumber \\
m_3^2 &=& \dfrac{\Lambda^{4}}{\mathcal{M}^2_{pl}} \left( Z \mathcal{F}_{Z} + 2 Z^2 \mathcal{F}_{ZZ1} \right) , \nonumber \\
m_4^2 &=& \dfrac{\Lambda^{4}}{\mathcal{M}^2_{pl}} \gamma \left[ Z \mathcal{F}_{Z} + 2 Z^2 \left( 3 \mathcal{F}_{ZZ1} + 2 \mathcal{F}_{ZZ2} \right) \right] ,
\end{eqnarray*}
where
\begin{eqnarray*}
\dfrac{\partial \mathcal{F}}{\partial Z^{ij}} &\equiv& \mathcal{F}_{Z} \delta_{ij}, \\
\dfrac{\partial^2 \mathcal{F}}{\partial Z^{ij} Z^{kl}} &\equiv& \mathcal{F}_{ZZ1} \delta_{ij} \delta_{kl} + \mathcal{F}_{ZZ2} \left( \delta_{ik} \delta_{jl} + \delta_{il} \delta_{jk} \right) .
\end{eqnarray*}
For this particular case, $- 3 \gamma p_\phi = \rho_\phi$ and the
following relations hold
\begin{eqnarray*}
\left\{ \begin{array}{lcl}
m_0^2 &=& 3 \gamma \left( m_4^2 - m_1^2 / 2 \right) , \\
m_4^2 &=& \gamma \left( 3 m_3^2 - m_2^2 \right) , \\
m_1^2 &=& 2 \left( 3 \gamma - 1 \right) p_\phi .
\end{array} \right.
\end{eqnarray*}

\section{The linearized equations}
\label{sec:linearized-equations-1}

Small perturbations on the cosmological background can be expressed
trough eq.~(\ref{metric-pert}-\ref{Gold-pert}). The tensor field
$h_{ij}$ and the vector field $\zeta_i$ are gauge-invariant, while the
other two gauge-invariant vector fields and 
seven gauge-invariant scalar fields are introduced trough
eq.~(\ref{gauge-vector}-\ref{gauge-scalars}). With all these
definitions, the linearized part of the Einstein tensor is given by
\begin{eqnarray*}
\delta \mathcal{G}_{00} &=& 2 \partial_i^2 \Psi - 6 \mathcal{H} \psi^\prime \label{A10} , \nonumber \\
\delta \mathcal{G}_{i0} &=& \partial_{i} \left[ 2 \left( \mathcal{H} \varphi + \psi^\prime \right) + \left( 2 \mathcal{H}^{\prime} + \mathcal{H}^2 \right) B \right] + \frac{1}{2} \partial_i^2 \varpi_{i} \nonumber \\
& & + v_i \left( 2 \mathcal{H}^{\prime} + \mathcal{H}^2 \right) \label{A11} , \nonumber \\
\delta \mathcal{G}_{ij} &=& - \delta_{ij} \left[ 2 \left( \varphi + \psi \right) \left( 2 \mathcal{H}^2 + \mathcal{H}^\prime \right) - 2 \mathcal{H} \left( 2 \psi + \varphi \right)^\prime \right. \nonumber \\
& & \left. - 2 \psi^{\prime\prime} + \partial_i^2 \left( \Psi - \Phi \right) - 6 \left( \psi + \varphi \right) \left( \mathcal{H}^2 + \mathcal{H}^\prime \right) \right] \nonumber \\
& & + \partial_i \partial_j \left[ \Psi - \Phi - 2 \left( 2 \mathcal{H}^{\prime} + \mathcal{H}^2 \right) E \right] - \mathcal{H} h_{ij}^{\prime} \nonumber \\
& & + \frac{1}{2} \left( \partial_i^2 h_{ij} - h_{ij}^{\prime\prime} \right) + \left( 2 \mathcal{H}^{\prime} + \mathcal{H}^2 \right) h_{ij} \nonumber \\
& & + \mathcal{H} \left( \partial_i \varpi_j + \partial_j \varpi_i \right) + \frac{1}{2} \left( \partial_i \varpi^\prime_j + \partial_j \varpi^\prime_i \right) \nonumber \\
& & - \left( \mathcal{H}^2 + 2 \mathcal{H}^\prime \right) \left( \partial_i F_j + \partial_j F_i \right) \label{A12} .
\end{eqnarray*}
The linearized energy-momentum tensor of the ordinary
matter fields is given by eq.~(\ref{energy-tensor-pert}), while the
linearized energy momentum tensor of the Goldstone fields takes the
form
\begin{eqnarray*}
\delta \mathcal{T}_{00}^{\phi} &=& a^2 M^2_{Pl} \left[ \left( 2 \rho_\phi + 2 \rho_\Lambda - m_0^2 \right) \varphi + m_0^2 \xi^{0\prime} / \phi^{\prime} \right. \nonumber \\
& & \left. + \left( \rho_\phi + p_\phi + m_4^2 \right) \left( \partial_i^2 \Xi + 3 \psi \right) \right] \label{A17} , \\
\delta \mathcal{T}_{0i}^{\phi} &=& \left[ \frac{\rho_\phi + p_\phi}{M^2_{Pl}} \left( \xi_i + \partial_i \xi \right)^{\prime} + \dfrac{\rho_\phi + \rho_\Lambda}{M^2_{Pl}} \left( v_i + \partial_i B \right) \right. \nonumber \\
& & \left. + \frac{m_1^2}{2} \left( v_i + \partial_i \mathcal{B} + \xi_i^{\prime} + \partial_i \Xi^{\prime} \right) \right] a^2 M^2_{Pl} \label{A18} , \\
\delta \mathcal{T}_{ij}^{\phi} &=& a^2 M^2_{Pl} \left[ \frac{1}{2} m_2^2 \left( \partial_j \xi_i + \partial_i \xi_j + 2 \partial_i \partial_j \xi \right) \right. \nonumber \\
& & - m_3^2 \delta_{ij} \left( 3 \psi + \partial_i^2 \Xi \right) + \delta_{ij} m_4^2 \left( \varphi - \xi^{0\prime} / \phi^{\prime} \right) \nonumber \\
& & + \left( \frac{1}{2} m_2^2 - \dfrac{p_\phi + p_\Lambda}{M^2_{Pl}} \right) \left( 2 \psi \delta_{ij} - \partial_i F_j - \partial_j F_i \right. \nonumber \\
& & \left. \left. - 2 \partial_i \partial_j E + h_{ij} \right) \right] \label{A19} .
\end{eqnarray*}

These relations allow to write the linearized Einstein field equations
for the massive gravitational field. They consist of one tensor
equation
\begin{eqnarray*}
0 &=& h_{ij}^{\prime\prime} - \partial_i^2 h_{ij} + 2 \mathcal{H} h_{ij}^{\prime} + a^2 m_2^2 h_{ij} ,
\end{eqnarray*}
two vector equations
\begin{eqnarray*}
0 &=& \varpi^\prime_i + 2 \mathcal{H} \varpi_i - a^2 m_2^2 \sigma_i ,
\end{eqnarray*}
\begin{eqnarray*}
0 &=& a^{-2} \partial_i^2 \varpi_{i} - \left( m_1^2 + 2 \frac{\rho_\phi + p_\phi}{M^2_{Pl}} \right) \left( \varpi_i + \sigma_i^\prime \right) \nonumber \\
& & - 2 \frac{\rho_m}{M_{Pl}^2} \left( 1 + \omega_m \right) \zeta_i ,
\end{eqnarray*}
and four scalar equations
\begin{eqnarray*}
0 &=& - 2 \partial_j^2 \Psi + 6 \mathcal{H} \left( \mathcal{H} \Phi + \Psi^\prime \right) - a^2 m_0^2 \left[ \frac{\phi^{\prime\prime}}{\phi^{\prime}} \mathcal{B} + \mathcal{B}^{\prime} + \Phi \right] \\
& & + a^2 \left( \frac{\rho_\phi + p_\phi}{M_{Pl}^2} + m_4^2 \right) \left( \partial_j^2 \Xi + 3 \Psi \right) + a^2 \frac{\rho_m}{M_{Pl}^2} \delta_\rho \nonumber , \\
0 &=& 2 \partial_i \left( \Psi^\prime + \mathcal{H} \Phi \right) + a^2 \partial_i \left[ - \frac{m_1^2}{2} \left( \mathcal{B} + \Xi^\prime \right) \right. \nonumber \\
& & \left. - \frac{\rho_\phi + p_\phi}{M^2_{Pl}} \Xi^\prime - \frac{\rho_m}{M_{Pl}^2} \left( 1 + w_\omega \right) \delta_\zeta \right] , \\
0 &=& \partial_i \partial_j \left( \Phi - \Psi + a^2  m_2^2 \Xi \right) , \\
0 &=& - 2 \Psi^{\prime\prime} - 2 \Phi \left( \mathcal{H}^2 + 2 \mathcal{H}^\prime \right) + \partial_j^2 \left( \Psi - \Phi \right) \nonumber \\
& & - 2 \mathcal{H} \left( 2 \Psi + \Phi \right)^\prime + a^2 \frac{p_m}{M_{Pl}^2} \delta_p - a^2 m_3^2 \partial_j^2 \Xi \nonumber \\
& & + a^2 m_4^2 \left( \Phi + \frac{\phi^{\prime\prime}}{\phi^{\prime}} \mathcal{B} + \mathcal{B}^{\prime} \right) + a^2 \left( m_2^2 - 3 m_3^2 \right) \Psi .
\end{eqnarray*}
The linearized Goldstone equations give one vector equation
\begin{eqnarray*}
0 &=& a^{-4} \partial_0 \left[ a^4 \left( m_1^2 + 2 \frac{\rho_\phi + p_\phi}{M^2_{Pl}} \right) \left( \varpi_i + \sigma_i^\prime \right) \right] - m_2^2 \partial_j^2 \sigma_i ,
\end{eqnarray*}
and two scalar equations
\begin{eqnarray*}
0 &=& \partial_0 \left[ \frac{a^4}{\phi^{\prime}} \left( m_0^2 \left( \frac{\phi^{\prime\prime}}{\phi^{\prime}} \mathcal{B} + \mathcal{B}^\prime + \Phi \right) - m_4^2 \left( 3 \Psi + \partial_i^2 \Xi \right) \right) \right] \nonumber \\
& & + \frac{a^4}{2 \phi^{\prime}} m_1^2 \partial_i^2 \left( \mathcal{B} + \Xi^\prime \right) , \\
0 &=& \partial_i \partial_0 \left\lbrace a^4 \left[ 
\left( \frac{m_1^2}{2} + \frac{\rho_\phi}{M_{Pl}^2} \right) 
(\mathcal{B} + \Xi') +  \frac{ p_\phi}{M^2_{Pl}}  
\Xi' \right] \right\rbrace \nonumber \\
& & + a^{4} \partial_i \left[ \left( m_3^2 - m_2^2 \right) \partial_j^2 \Xi - \left( m_4^2 + \dfrac{\rho_\phi + p_\phi}{M_{Pl}^2} \right) \Phi \right. \nonumber \\
& & \left. + \left( 3 m_3^2 - m_2^2 \right) \Psi - \left( m_4^2 + \frac{\rho_\phi}{M_{Pl}^2} \right) \left( \frac{\phi^{\prime\prime}}{\phi^{\prime}} \mathcal{B} + \mathcal{B}^\prime \right) \right] .
\end{eqnarray*}

\end{document}